# Impact dynamics of flexible hydrogels on solid substrates of different wettability

Akash Chowdhury[a], Surjyasish Mitra[a], and Sushanta K. Mitra[*,a,b]

[a]Micro & Nano-Scale Transport Laboratory, Waterloo Institute for Nanotechnology, Department of Mechanical and Mechatronics Engineering, University of Waterloo, 200 University Avenue West, Waterloo, Ontario N2L 3G1, Canada

[b]International School of Engineering (ISE), Chulalongkorn University, Bangkok, 10330, Thailand

*To whom correspondence should be addressed: Sushanta K. Mitra., E-mail: skmitra@uwaterloo.ca, Telephone: 519-888-4567, ext.37176

**Abstract:**
**Hypothesis**
The impact dynamics of Newtonian liquid drops and rigid spheres are extensively studied and are well understood. However, bridging the two necessitates the investigation of the impact of soft elastic spheres. We hypothesize that in such a scenario, the interplay of impact kinetic energy, elastic energy, and the work of adhesion governs the transition from liquid-like to solid-like impact behaviors.

**Experiments**
In this work, we perform experiments with spherical polyacrylamide (PAAm) hydrogel drops/spheres, spanning a broad range of shear moduli and impact velocities on hydrophilic (plasma-treated glass) and hydrophobic (silane-coated) substrates, yielding an elastic number $El$ variation of five orders of magnitude. Transient spreading morphology and impact force were simultaneously resolved using synchronized high-speed imaging and piezoelectric force sensing.

**Findings**
At low elastic numbers ($El < 1$), impacting hydrogels exhibit a hybrid poroelastic response: a liquid-rich contact foot is expelled from the polymer network and spreads independently, while the bulk drop undergoes viscoelastic contact-line pinning into a pancake geometry at maximum deformation. At high elastic numbers ($El > 1$), contact foot spreading is suppressed, and deformation is accurately described by a neo-Hookean energy balance, yielding a maximum spreading factor independent of substrate wettability. Further, we show that the normalized peak impact force $F^*$ collapses to a constant value consistent with the Wagner limit for $El < 1$ and follows a power-law scaling $F^* \sim El^{0.38}$ for $El > 1$, in close agreement with both Hertzian and neo-Hookean predictions, and independent of substrate wettability. Furthermore, we highlight that post-impact retraction is suppressed across nearly the entire parameter space due to adsorbed polymer chains anchoring the receding gel network to the substrate, producing circumferential ridge instabilities; rebound occurs only when elastic restoring forces overcome the work of adhesion.

## 1. Introduction

The precise deposition of soft material droplets onto substrates is central to modern additive manufacturing, particularly in the rapidly expanding field of hydrogel-based 3D bioprinting.[1–3] In bioprinting workflows, hydrogel bioinks, typically polyacrylamide or gelatin-based formulations, are dispensed as discrete droplets or continuous filaments onto build platforms at controlled velocities,[4–6] where their post-impact deformation directly governs print resolution, layer fidelity, and structural integrity of the resulting construct.[7–9] Despite this central role, the mechanics governing how a soft elastic sphere deforms upon impact on a substrate remain poorly understood, representing a critical knowledge gap between material formulation and print quality control.[10,11]

For simple Newtonian liquid droplets, the impact dynamics are well described by the Weber number (We) framework, whose maximum spreading factor $\beta$ scales as $We^{0.5}$, where $\beta$ is defined as the ratio of the maximum diameter of the deformed droplet to its initial diameter.[12] Here, $We = (\rho v_0^2 r_0)/\gamma$, where $\rho$ and $\gamma$ are the liquid density and surface tension, respectively, $v_0$ is the impact velocity, and $r_0$ is the initial radius of the drop. However, polymer-based bioinks occupy an intermediate regime between a viscous fluid and an elastic solid.[13,14] Their rheological character, tunable through monomer concentration, indicates that a bioink droplet may spread aggressively and pin irreversibly on impact, or recoil elastically and partially detach, both of which are detrimental outcomes in a layer-by-layer fabrication context. Spreading beyond the intended voxel footprint for 3D bioprinting degrades lateral resolution, while elastic rebound disrupts adhesion between successive layers, compromising the mechanical continuity of printed scaffolds.[15–19]

Preliminary work by Tanaka *et al.* revealed that for the impact of soft gels, instead of the Weber number dependence, the maximum spreading factor scales as $\beta \sim Ma$, where $Ma$ is the Mach number defined as $v_0/\sqrt{G/\rho}$.[10,11] Here, $G$ is the modulus of rigidity (shear modulus) of the gel and $\sqrt{G/\rho}$ represents the velocity of transverse sound waves in an elastic medium. Further refinement by Arora *et al.* introduced a modified Mach number to interpret the impact dynamics, i.e., $Ma^* = v_0/\sqrt{2((G/\rho) + (3\gamma/\rho d_0))}$, where $\gamma$ is the surface tension of the gel and $d_0$ is the initial diameter of the gel drop, to incorporate surface energy effects via the surface tension.[20] Though this approach aimed to provide a unified description of soft impact across a broad range of elastic modulus, their experiments employed non-contact Leidenfrost substrates, deliberately eliminating wettability effects. Thus, the role of substrate wettability in modulating post-impact contact propagation, gel deformation, and ultimately contact suppression has not been systematically addressed in the context of soft gel impact.

Equally critical, yet largely overlooked in existing literature, is the peak impact force transmitted to the substrate upon impact.[21] Recent works by Lohse and coworkers have shown that for conventional liquid drop impact on rigid surfaces, the peak force scales as $F_{\max} \sim \rho v_0^2 r_0^2$, at par with Wagner's theory.[22–24] Additionally, they highlighted that the impact of elastic spheres is governed by Hertz theory, which introduces a modulus-dependent peak force scaling, $F_{max} \sim G^{0.4}(\rho v_0^2)^{0.6} r_0^2$ which can be rewritten as $F_{\max} \sim El^{0.4} \rho v_0^2 r_0^2$ where $El = G/\rho v_0^2$ is the elastic number.[25] Note that this scaling law is theoretically proposed and remains to be validated experimentally.[25] It is noteworthy to mention here that, for bioprinting onto delicate substrates, such as pre-deposited cell-laden hydrogel layers, functionalized membranes, or compliant tissue scaffolds, these transient impact forces can induce local structural damage and delamination, even when the final printed geometry appears acceptable.[26,27] Providing reliable predictive estimates of these forces as a function of ink (gel) elasticity and print velocity is therefore of direct practical relevance to the field.

In this work, we directly confront these open questions using polyacrylamide hydrogels across a broad range of elastic modulus, spanning the liquid-to-solid transition, impacted at varying velocities onto both hydrophilic (plasma-treated glass) and hydrophobic (silane-coated) substrates. We simultaneously characterize the post-impact spreading morphology via high-speed imaging and the peak impact force via piezoelectric force sensing. Consequently, we aim to develop a predictive framework to understand post-impact deformation characteristics as well as validate previously proposed scaling laws that bridge the Wagner (fluid) and Hertz (elastic solid) limits.

## 2. Experimental

### 2.1. Fabrication of Hydrogels and Substrate

Polyacrylamide hydrogels were prepared using acrylamide (AAm) as the monomer and N,N′-Methylenebisacrylamide (BIS) as the cross-linker, and 2,4,6-trimethylbenzoyl-diphenylphosphine oxide (TPO) nanoparticle as the initiator. The detailed fabrication protocol is provided in our previous works.[13,28,29] Briefly, pregel solutions of 33.6 µl (~2 mm radius) were suspended in a beaker with n-octane and silicone oil (Silicone AP100, Sigma Aldrich), with a 1:2 volumetric ratio between the two. The density gradient between n-octane (density of 0.71 g/cm$^3$) and silicone oil (density of 1.08 g/cm$^3$) was achieved by first pouring heavy silicone oil and then slowly pouring octane on top of the oil. The bilayer solvent was kept for 3 hrs to ensure a smooth density gradient. 33.6 µL of hydrogel was pipetted into the middle of the solvent gradient and formed a spherical shape. This was cured under UV light (~365 nm) for 20 mins. The cured hydrogels were washed with hexane multiple times before each use. The properties of hydrogels are given in Table 1. The hydrogels are named PAAmX, where X denotes the acrylamide monomer ratio used to fabricate the hydrogel. Note that shear modulus was measured for cured hydrogel sheets using a rheometer, while the density and surface tension were measured for uncured hydrogel in its solvent state.

Table 1: Physical properties of PAAm hydrogels for different monomer weight percentages.

| Hydrogel | Shear Modulus, $G$ (kPa) | Surface Tension, $\gamma$ (mN/m) | Density, $\rho$ (kg/m$^3$) |
|---|---|---|---|
| PAAm6.3 | 0.05 | 65.7±0.1 | 1003.1 |
| PAAm6.5 | 0.06 | 65.6±0.2 | 1003.3 |
| PAAm7.5 | 0.2 | 64.8±0.3 | 1004.3 |
| PAAm10 | 10.24 | 64.2±0.2 | 1006.8 |
| PAAm13 | 35.55 | 61.9±0.1 | 1009.8 |
| PAAm20 | 97.61 | 60.6±0.2 | 1013.3 |
| PAAm30 | 130.94 | 57.3±0.1 | 1019.4 |

For glass substrates, microscope slides (76.2 × 25.4 × 1.0 mm, Bev-l-edge, Propper) were cleaned with isopropyl alcohol (IPA), DI water, and hexane, followed by plasma treatment for 2 minutes, making them hydrophilic. For silane-coated glass substrates, fresh microscope slides (76.2 mm × 25.4 mm × 1.0 mm, Bev-l-edge, Propper) were cleaned and plasma-treated similarly. Consequently, a millimeter-sized drop of (heptadecafluoro-1,1,2,2-tetrahydrodecyl)trichlorosilane (Sigma-Aldrich) was dispensed on a filter paper. The filter paper and cleaned microscope slides were kept in a desiccator, and a vacuum pump was switched on for 15 minutes and turned off. The slides and filter paper were kept in that condition inside the desiccator overnight (~12 hrs). The water contact angle on glass substrates was less than 10º, while on silane substrates it was 109º.

## 2.2. Experimental Setup

Figure 1 represents the experimental setup for conducting the impact experiments. The hydrogel spheres were held on the nozzle by suction and released gradually from a vertical height of $H$. In this study, $H$ were 5.1 cm, 20.4 cm, and 45.9 cm, corresponding to impact velocities ($v_o = \sqrt{2gH}$) of 1 m/s, 2 m/s, and 3 m/s, respectively. The corresponding Weber numbers (based on drop radius) are approximately 30, 120, and 270, with slight variation due to minor differences in the density and surface tension of the materials. The substrate was attached to a sensor mount, which was in turn connected to a piezoelectric force sensor (PCB 209C11) via threads. The sensor was fixed to the ground with an additional mount via threads and bolts, as shown in the figure. The force sensor has a sensitivity of 567 mV/N. The sensor was connected to a signal conditioner (PCB 482C05), and the output from the signal conditioner was captured by an oscilloscope (Siglent SDS1104X-E). The instantaneous spike generated from the force sensor upon impact was captured by the oscilloscope, which also served as the trigger for the high-speed camera (Photron S12) used to capture the impact dynamics at a frame rate of 10000 fps.

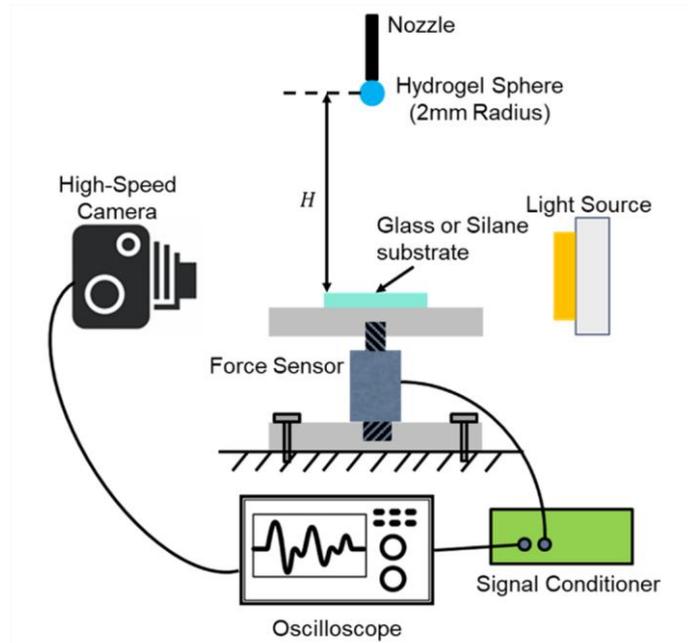

**Figure 1**: **Schematic of the experimental setup.** A hydrogel sphere ($r_0 = 2$ mm) is impacted from a vertical height, $H$, on a glass (water contact angle <10º) or silane substrate (water contact angle ~109º). The impact dynamics are recorded at 10,000 fps using a high-speed camera, and the impact forces are measured using a piezoelectric force sensor fixed to the bottom of the substrate.

## 3. Results and Discussion

Soft gels, upon impact, on a rigid substrate undergo a rapid deformation stage until achieving a maximum diameter, followed by a receding stage driven by the stored elastic energy, and depending on adhesion, may attempt to detach from the substrate. To characterize the deformation, we define the elastic number ($El = G/\rho v_0^2$), which scales the material's stored elastic energy against the impact kinetic energy. In the experimental range of velocities and moduli probed, the elastic number spans five orders of magnitude (0.006 to 128), dictating the transition of drop shapes at maximum deformation. As shown in Figure 2, the morphologies of impacting elastic drops at maximum deformation are consistent with regimes identified by Tanaka *et al.*: for weak elasticity, a concave pancake shape, an ellipsoid shape for intermediate elasticity, and a Hertzian shape for high elasticity.[11] The profile change can be observed with an increasing elastic number. The maximum lateral diameter ($d_{max}$) of the deformed shapes is used to define the maximum spreading factor: $\beta = d_{max}/2r_0$ (see Figure 2).

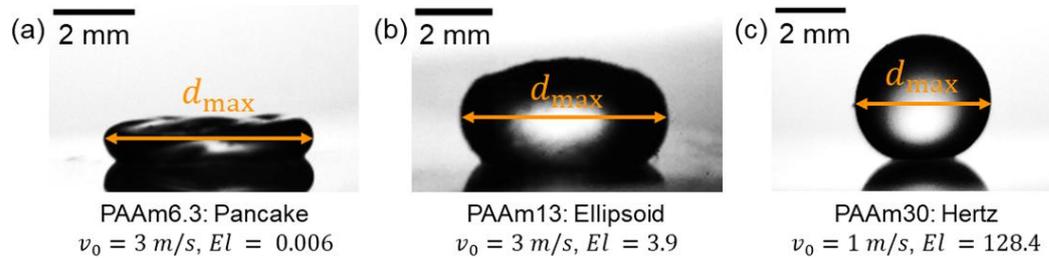

Figure 2: **Configuration of hydrogels (initial radius $r_0 \approx 2$ mm) with varying elasticity at maximum deformation after impact.** (a) Hydrogel with a lower elastic number (PAAm6.3 with $G = 0.05$ kPa, impacting at $v_0 = 3$ m/s, $El = 0.006$) deforms into a pancake shape. (b) Hydrogel with a moderate elastic number (PAAm13 with $G = 35.55$ kPa, impacting at $v_0 = 3$ m/s, $El = 3.9$) deforms into an ellipsoid shape. (c) Hydrogel with a high elastic number (PAAm30 with $G = 130.94$ kPa impacting at $v_0 = 1$ m/s, $El = 128.4$), undergoes a local Hertzian deformation. Scale bars represent 2 mm.

### 3.1. Morphology of spreading

Figure 3 illustrates the post-impact contact dynamics across an elastic number gradient from $El = 0.007 - 10.7$ as we vary the PAAm's shear modulus from $G = 0.06 - 90.7$ kPa, highlighting a transition in impact morphology with gel elasticity. In our previous work, we have reported the presence of a contact foot, also referred to as the deformation foot, in hydrogels whose height varies with $E^{-1}$.[13] The contact foot mostly consists of the liquid attracted out of the gel network towards the surface.[30,31] For softer hydrogels, the contact foot is larger, and thus, during spreading, a fraction of the contact foot is ejected and spreads outward like a liquid upon impact, as seen in PAAm6.5 ($G = 0.06$ kPa, $El = 0.007$) and PAAm7.5 ($G = 90.7$ kPa, $El = 0.049$) (see Figure 3). Notably, this expelled liquid remains separated from the main drop and does not reabsorb into the drop. Meanwhile, the bulk of the drop undergoes a pinning process and forms a pancake shape (the pinning process is further elaborated in the next figure).

As the drop elasticity increases, the lateral extent of the deformation foot diminishes. For PAAm10 ($G = 10.24$ kPa, $El = 2.54$), we observe that the foot spreads only slightly before being absorbed back into the bulk drop, and for PAAm13 ($G = 35.55$ kPa, $El = 3.91$) onwards, the emergence of the deformation foot is entirely suppressed (see Figure 3). This shift indicates a fundamental transition in impact physics. Standard impact dynamics for soft spherical solids are typically described by the ellipsoid and Hertzian regimes governed by Hookean elasticity,[11,32] whereas liquid drop impact is characterized by large, thin sheets controlled by a balance of viscous dissipation and surface tension.[12] Our findings demonstrate that soft hydrogels occupy a unique intermediate state between these two extremes. These materials exhibit a hybrid response, where a small fraction of the mass undergoes liquid-like spreading, while the bulk drop experiences viscoelastic pinning and stores impact energy as elastic strain energy within its pancake-like geometry.

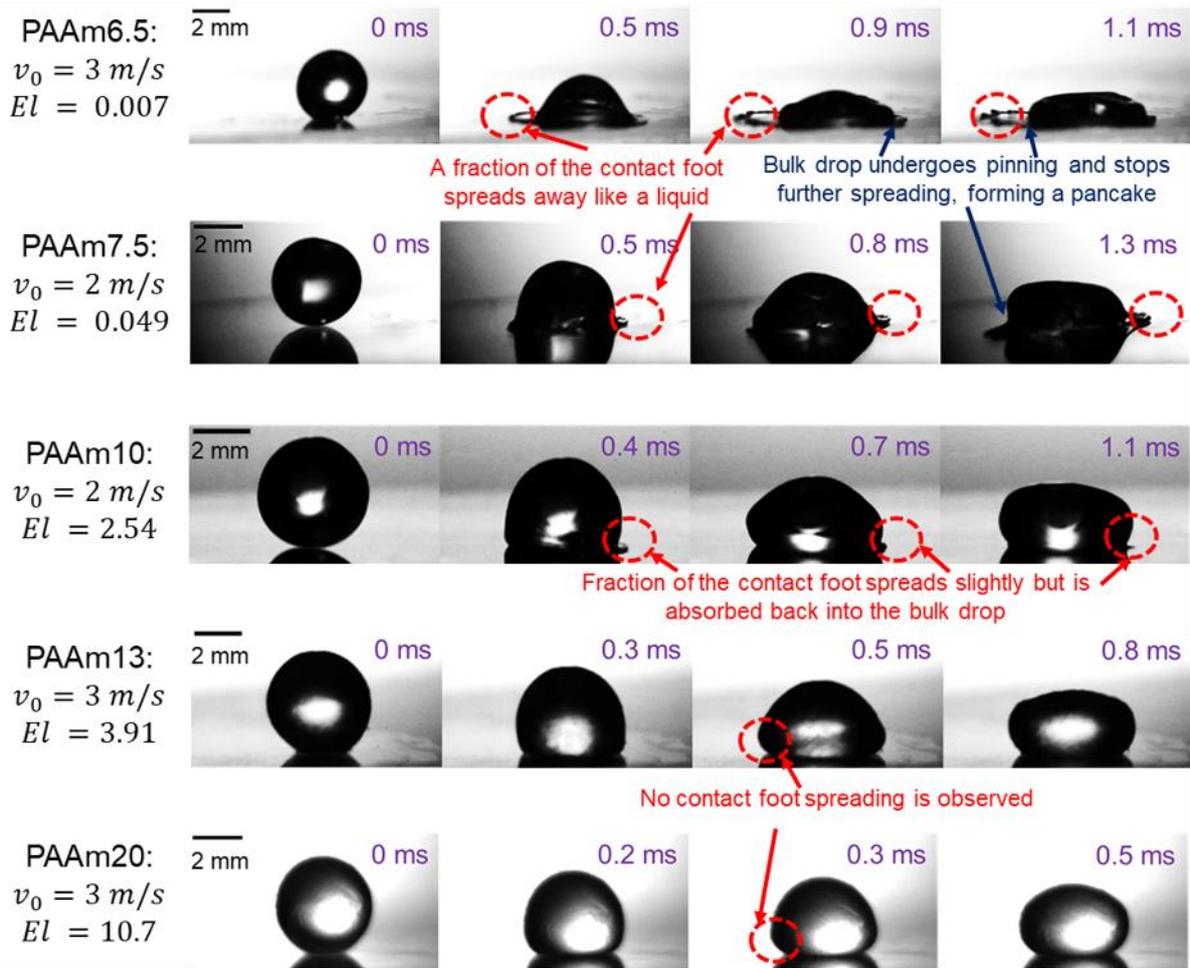

**Figure 3**: **Experimental snapshots highlighting the post-impact contact/spreading dynamics and morphological transitions for hydrogels with varying elasticity or conversely with varying elastic number ($El = 0.007 - 10.7$).** The observed morphological transition is driven by the behavior of the contact foot of the hydrogel. For ultra-soft hydrogels (PAAm6.5 and PAAm7.5), a significant fraction of this foot is ejected and spreads independently like a fluid, while the bulk drop undergoes viscoelastic pinning to form a stable pancake. As elasticity increases (PAAm10), the spreading of the foot decreases and is reabsorbed into the bulk. In the stiffer hydrogels (PAAm13 and PAAm20), liquid foot spreading is entirely suppressed, and the drop achieves an ellipsoid shape through purely elastic deformation. Scale bars represent 2 mm.

Multiple impacts with the same PAAm6.5 drop ($G = 0.06$ kPa, $El = 0.007$) were performed, as shown in Figure 4. The drop after multiple impacts undergoes a slight permanent deformation and has a non-spherical shape at $t = 0$ ms (see Figure 4(b)). The portion of the liquid foot that spreads in a fresh drop (Figure 4(a)) is not replenished. Thus, the liquid spreading is not observed after multiple impacts. However, both fresh and previously impacted drops exhibit consistent contact line pinning and pancake formation. An image sequence explaining the spreading morphology is shown in Figure 4(c). As the hydrogel spreads, the contact line expands slowly, and an outward bulge forms, directing flow away from the contact line. The contact foot is made of liquid water with dissolved small chains of PAAm, essentially a viscoelastic liquid. In the spreading of viscoelastic liquid, the formation of a dissipative sink near the contact line, leading to the pinning of the drop, is widely reported.[33–35] The contact line is pinned at 0.9 ms, and the remaining vertical inertia is translated into a bulge outward from the contact line, which eventually forms a pancake shape, during which the contact area remains unchanged. Hence, for subsequent maximum spreading factor calculations, we consider the diameter of this pancake shape. While this dual response is evident, a detailed analysis of the fractional liquid spreading remains beyond the scope of the current work.

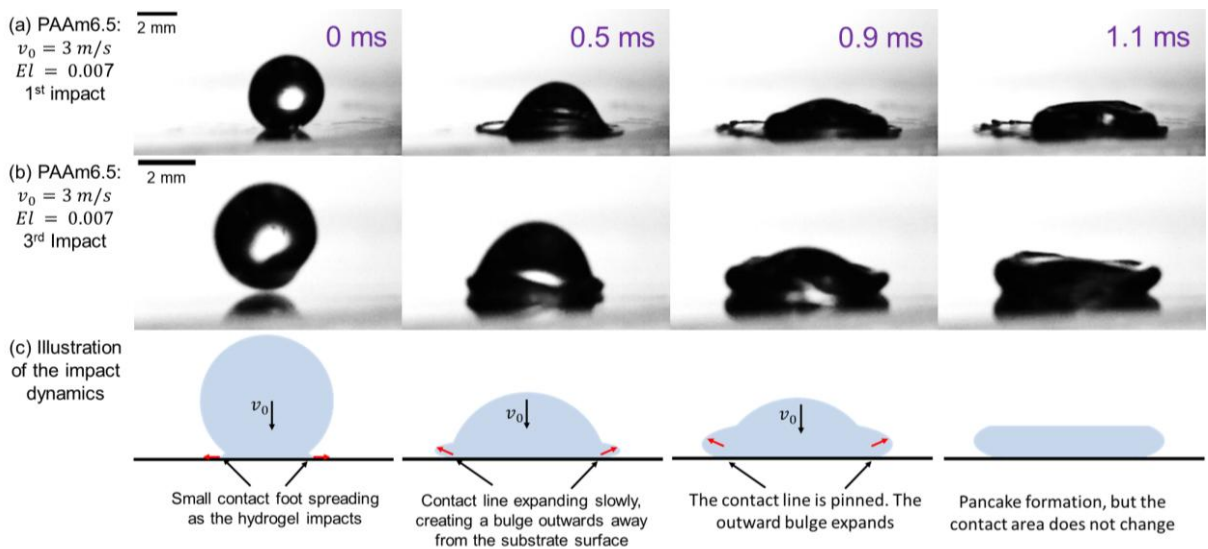

**Figure 4: Analysis of multiple impacts of the same drop of PAAm6.5 hydrogel.** (a) Experimental snapshots showing the initial impact (1$^{st}$ impact) of a fresh drop, highlighting the ejection and liquid-like spreading of a fraction of the contact foot. (b) Experimental snapshots showing the third impact of the same PAAm drop, demonstrating that the contact foot is not replenished after the first event, resulting in a lack of liquid-like spreading. (c) Schematic illustration of the impact morphology of the elastic PAAm drop after 3$^{rd}$ impacts: initial contact foot spreading is followed by a slow expansion of the contact line, creating an outward bulge. The contact line becomes pinned (at approximately 0.9 ms after impact), directing the remaining vertical inertia into the expanding bulge to form a stable pancake geometry while the contact area remains unchanged. Scale bars represent 2 mm.

### 3.2. Maximum spreading diameter

To elaborate on the transient contact morphology for the impact of elastic drops and how they differ from the solid/liquid limit, here we analyze the maximum spreading diameter. The elastic strain energy $E_g$ stored at the moment of maximum deformation can be estimated using the neo-Hookean model, as $E_g = 2/3\pi r_0^3 G(2\beta^2 + \beta^{-4} - 3)$.[20] Existing literature investigating the impact of weakly elastic polymeric liquids typically assumes large deformations, i.e., $\beta \gg 1$, and approximates $E_g \approx 4\pi r_0^3 G\beta^2/3$. However, for the intermediate elasticity probed in the present study, such simplifications are not valid (see Note S1 and Figure S1, Supplementary Information). In our experiments with PAAm shear modulus varying between 0.05 kPa – 97.61 kPa, and impact We number varying between 29 – 320, we observed $1.02 < \beta < 1.69$. Thus, we consider the full expression of the elastic strain energy for subsequent analysis. Assuming the drop transforms into a pancake of diameter $\beta r_0$ at maximum deformation, as shown in Figure 5(a), the energy balance can be expressed as:

$$T = E_g + E_s - W_a + L, \qquad (1)$$

where $T$ is the initial kinetic energy, $E_s$ is the surface energy contribution from the top surface and the curved surface of the pancake, $-W_a$ is the interfacial energy contribution from the bottom contacted surface of the pancake, involving the work of adhesion $w \approx 2\gamma$, and $L$ is the dissipative losses. Consequently, Eq. (1) can be written as,

$$\frac{2}{3}\pi r_0^3 \rho v_0^2 = \frac{2}{3}\pi r_0^3 G(2\beta^2 + \frac{1}{\beta^4} - 3) + \pi\gamma\beta^2 r_0^2 + \frac{\pi\gamma r_0^2}{\beta} - \pi w\beta^2 r_0^2 + L \qquad (2)$$

Normalizing Eq. (2) w.r.t. $T$, we get:

$$1 - l = El\left(2\beta^2 + \frac{1}{\beta^4} - 3\right) - \frac{3}{2We}\beta^2 + \frac{4}{We\beta}, \qquad (3)$$

where $l = L/T$ is the fraction of kinetic energy lost in dissipation. Eq. (3) can be rearranged as a 6-order polynomial in $\beta$ as follows:

$$\beta^6 - \frac{1.5 + \frac{0.5}{El}(1-l)}{1 - \frac{0.75}{We*El}}\beta^4 + \frac{4}{2We*El - 1.5}\beta^3 + \frac{0.5}{1 - \frac{0.75}{We*El}} = 0 \qquad (4)$$

Eq. (4) has important consequences. For instance, in the rigid limit, i.e., $El \to \infty$, the solution of Eq. (4) yields -0.5, 1, 1 as roots, validating zero deformation for rigid bodies. The expression $We * El = Gr_0/\gamma = Ec$ is the elastocapillary number, which varies between 1.2 – 4433 in the present experiments and is independent of the impact velocity. Thus, the maximum spreading factor $\beta = f(El, Ec, l)$ is primarily dependent on the elastic number for each hydrogel. Further, for stiffer hydrogels (PAAm10 and higher), $Ec >> 100$, and the maximum spreading factor has weak dependence on the elastocapillary number and converges to $\beta = f(El, Ec \to \infty, l)$. Considering zero losses ($l = 0$), Eq. (4) can be reduced to:

$$(\beta^2)^3 - (1.5 + \frac{0.5}{El})(\beta^2)^2 + 0.5 = 0 \qquad (5)$$

Eq. (5) is a cubic equation in $\beta^2$ and can be solved analytically. In Figures 5(b) and 5(c), we demonstrate the variation of the experimentally obtained $\beta$ across five decades of elastic numbers for impact on silane and glass substrates. The blue dashed line, representing $\beta = f(El, Ec \to \infty, 0)$, is in good agreement with the experimentally obtained data for $El > 1$ for both the substrates.

For softer hydrogels ($El < 1$), the experimental data indicates a significantly lower spread compared to the solution of Eq. (5). Hence, for softer gels, numerous dissipation mechanisms and surface effects dominate the phenomena. The experimental data is observed to lie between the numerical solution of Eq. (4) for $Ec = 6.1$ and $l = 0.9$, and $Ec = 1.2$ and $l = 0.99$, as shown with purple and red dashed lines, respectively. This observation signifies that a substantial portion (90% to 99%) of the drop's initial kinetic energy is diverted into several viscous and viscoelastic dissipation modes, including the poroelastic squeeze and spreading of water contact foot[36,37], with a possibility of micro network damage. More importantly, the phenomenon, often referred to as the 'dissipative sink', occurs near the three-phase contact line for viscoelastic liquids. As the gel deforms, the energy required to move the contact line increases, effectively "pinning" it and prematurely halting the spreading process, as shown in Figure 4 in the previous section.[33–35]

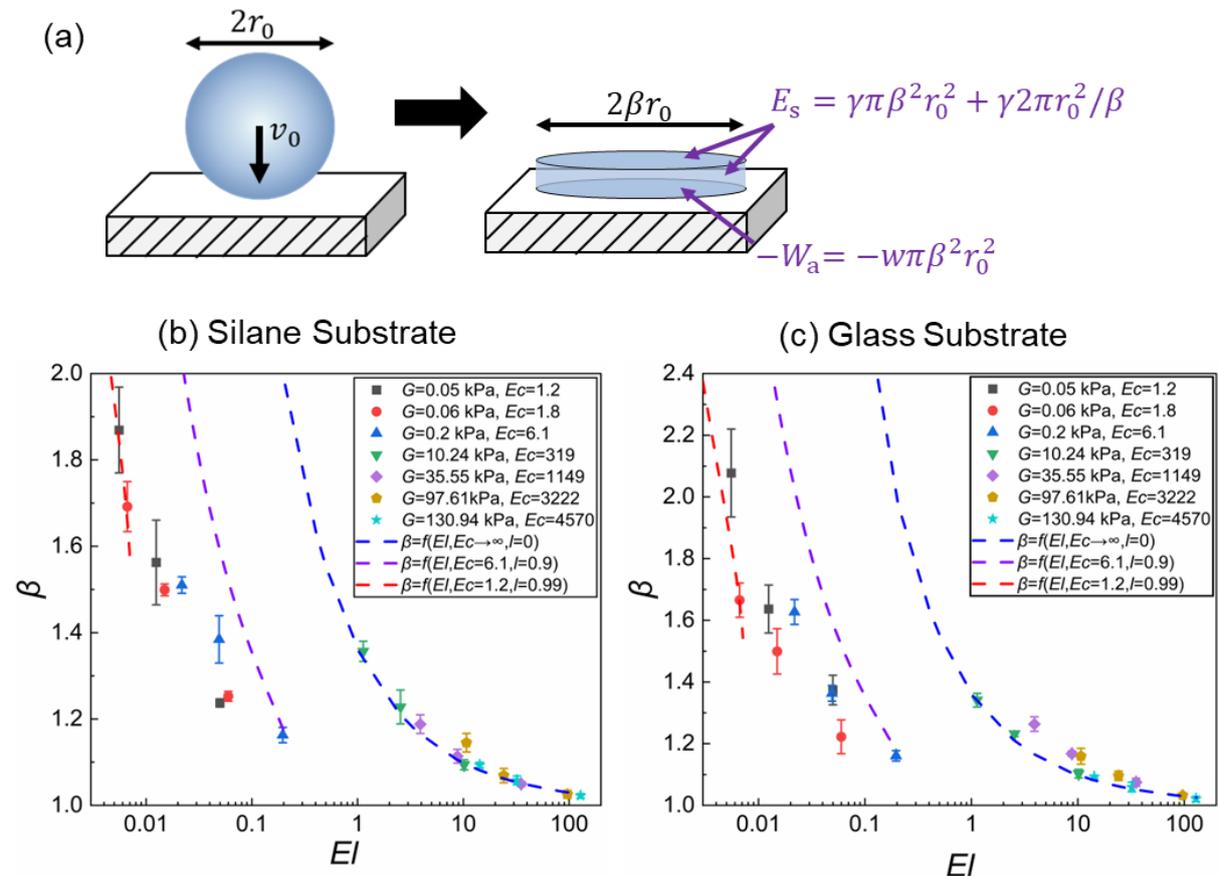

**Figure 5: Scaling of the maximum spreading factor, $\beta$ with elastic numbers, $El$, highlighting the crossover from inertial-viscous to elasticity-dominated impact regimes.** (a) Impact Schematic: A sphere of radius $r_0$ impacting the surface at a velocity $v_0$ and deforming into a pancake shape with a new radius $\beta r_0$. The contributions from the surface energy and work of adhesion are explicitly shown. (b,c) Experimental plot showing the variation of the maximum spreading factor $\beta$ with the Elastic number $El$ for PAAm hydrogel drops of varying elasticity impacting on silane-treated (b) and glass (c) substrates. The blue dashed lines represent the theoretical model $\beta = f(El, Ec \to \infty, l = 0)$ which satisfies for $El > 1$ cases. While the spreading factor is lower in softer hydrogels and lies between $\beta = f(El, Ec = 6.1, l = 0.9)$ and $\beta = f(El, Ec = 1.2, l = 0.99)$ represented by the purple and red dashed lines, respectively.

### 3.3. Maximum Impact Force

The impact force exerted by the elastic drop on the substrate typically peaks at maximum deformation before decreasing during the receding stage. The normalized transient impact force recorded by the force sensor is shown in Figures 6(a) and 6(b) for two extreme elastic numbers: PAAm7.5 ($G = 0.2$, $El = 0.049$) and PAAm30 ($G = 130.94$, $El = 128$), respectively. The corresponding drop morphology at various instants is shown by the snapshots in the inset (Figure 6). From Figure 6(b), it is evident that at a high elastic number, the force response exhibits elastic characteristics; the spreading and receding stages are nearly in phase with the impact force, with the peak force occurring at the instant of maximum PAAm deformation and a force trough at the instant of recoil. The response is largely symmetric, with comparable durations of the spreading (0.44 ms) and receding (0.532 ms) stages. Notably, negative force values were recorded during receding, indicating an adhesive pull exerted by the drop on the substrate. Conversely, at a low elastic number, the response is skewed, with the receding stage (2.3 ms) being more than twice the spreading stage (1.1 ms) (Figure 6(a)). Further, the impact force remains positive even at maximum recoil, implying a lack of significant elastic recovery. Furthermore, the peak impact force is also an order of magnitude lower. Collectively, the force profile indicates a transition from liquid-like to elastic-like behavior with the increase in elastic number.

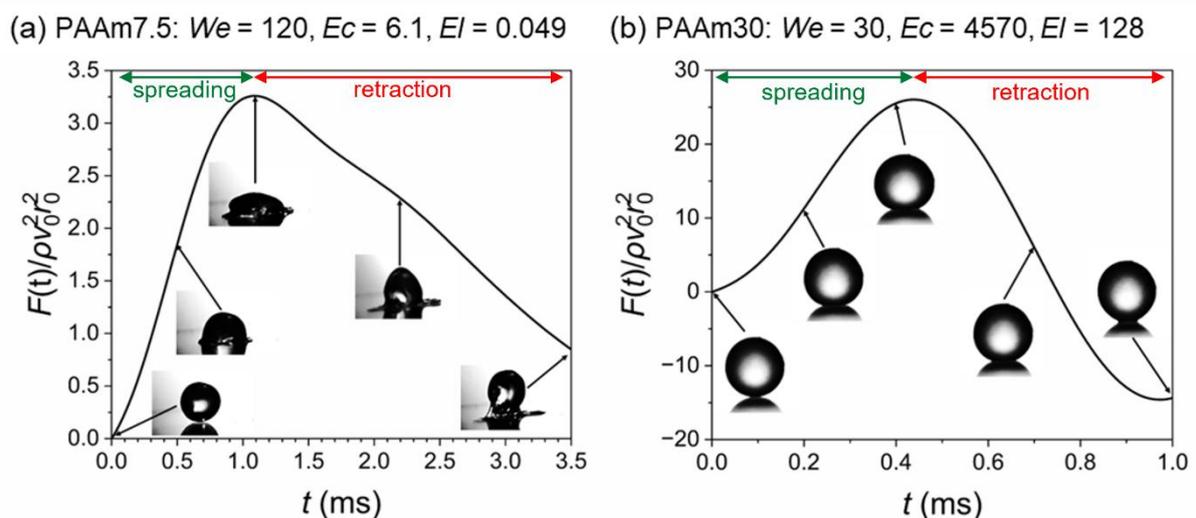

**Figure 6. Evolution of normalized impact force $F(t)/\rho v_0^2 r_0^2$ for impacts of PAAm drops on glass substrates.** (a) Impact of PAAm7.5 ($El = 0.049$), representing a low elastic number case, and (b) impact of PAAm30 ($El = 128$), representing a high elastic number case. Inset images show the corresponding drop morphology at various instants during impact. Note that for $El = 0.049$, the normalized peak force recorded is 3.28 at $t = 1.1$ ms, whereas for $El =$

128, the normalized peak force recorded is 26 at $t = 0.43$ ms. The post-impact spreading and retraction stages are marked for clarity.

We elucidate how the maximum impact force $F_{max}$ exerted by the impacting PAAm on the substrate varies with elastic properties. In the liquid limit, existing literature has revealed that Wagner's theory accurately describes the pressure distribution and force evolution for impacting drops. Under this framework, the maximum impact force or the peak force scales as $F_{max} \sim \rho v_0^2 r_0^2$, or the normalized force, $F^* = F_{max}/\rho v_0^2 r_0^2 \approx 3.24$ for impact regimes with Weber numbers of 10 or higher.[22] On the other hand, in the solid limit, the interaction is governed by the material's bulk deformation rather than by fluid flow. To describe this, Jana *et al.* applied an energy balance based on kinetic energy and non-adhesive Hertz contact theory, which assumes purely elastic deformation between a sphere and a flat surface. Consequently, their theoretical model revealed $F^* = 5.3 El^{0.4}$.[25] Conversely, using the neo-Hookean expression for $E_g$ and defining the impact force as $F = \frac{dE_g}{d\delta}$, where indentation, $\delta = 2r_0(1 - 1/\beta^2)$, the maximum impact force can be expressed as:

$$F_{max} = \frac{dE_g}{d\beta}\frac{d\beta}{d\delta} = \frac{2}{3}\pi G r_0^2 \left(\beta^4 - \frac{1}{\beta^2}\right) \tag{6}$$

Normalizing Eq. (6) with $\rho v_0^2 r_0^2$, and using the solution of Eq. (5), $\beta = f(El, Ec \to \infty, l = 0) = g(El)$, Eq. (6) reduces to:

$$F^* = \frac{2}{3}\pi El \left(g(El)^4 - \frac{1}{(g(El))^2}\right) \tag{7}$$

In Figure 7, we show the variation of the experimentally obtained normalized peak force $F^*$ with the elastic number $El$ for impact on silane (Figure 7(a)) and glass (Figure 7(b)) substrates. We compare our experimental findings with the Hertz and Wagner predictions as well as the solution of Eq. (7). From Figure 7, we observe that the peak force scales as $F^* \sim El^{0.37}$, for $El > 1$ and was constant with $F^* \approx 3.65$ for $El < 1$ for impact on silane. Similarly, for impact on glass substrates, we observe $F^* \sim El^{0.38}$, for $El > 1$, and was constant at $F^* \approx 3.42$ for $El < 1$.

Thus, the peak impact force is independent of the surface properties.[22] Further, for both substrates, the peak forces for $El < 1$ are slightly higher than the Wagner limit of 3.24, as the hydrogels are viscoelastic, while the value 3.24 corresponds to the Newtonian limit.[24] The theoretical limits for Wagner and Hertz behaviors are represented by the green and blue dashed lines, respectively. While the curve for Eq. (7) slightly deviates from pure power law, it is well approximated by a power law scaling ($F^* \sim El^{0.42}$) over the experimental range of $1 < El < 130$, shown by the orange dashed line. Interestingly, the power law scaling obtained for both models was similar (0.4 in Hertz and 0.42 in neo-Hookean), which aligns well with the experimentally obtained exponent of 0.37 - 0.38. Given that the maximum spreading diameter is accurately predicted by the energy balance, this lower prefactor is likely a result of measurement challenges in capturing the instantaneous peak force. Due to the rapid nature of the impact and high-frequency signal characteristics, a low-pass filter was applied to mitigate electronic noise. Consequently, the recorded peak force represents a slightly attenuated value rather than the instantaneous theoretical maximum (see Note S2 and Figure S2, Supplementary Information, for additional details on force measurement). While the experimental pre-factor is influenced by this sensor-level smoothing, the more significant power-law dependence was successfully recovered, aligning closely with the predicted theoretical scaling. This confirms that while the absolute magnitude is filtered, the underlying physics governing the elastic-to-viscous transition remains accurately captured.

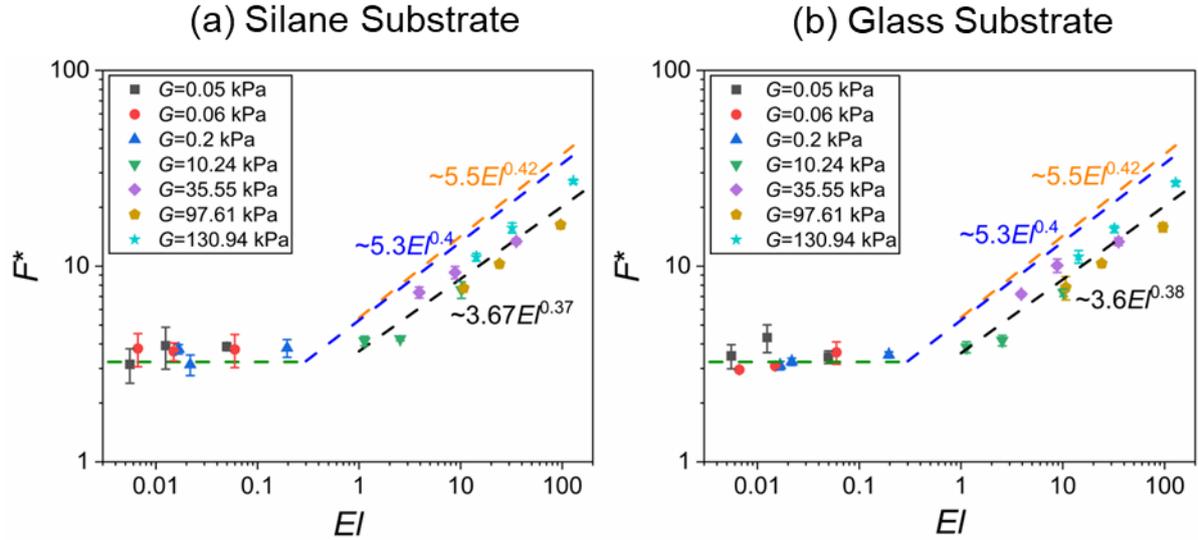

**Figure 7: Variation of normalized peak force, $F^* = F_{max}/\rho v_0^2 r_0^2$ with the elastic number $El$ for the impact of PAAm drops of different shear modulus $G$.** The peak force is independent of substrate wettability as seen in silane (a) and glass (b) substrates. All Weber numbers are incorporated within $El$. For $El > 1$, the data collapses into a single curve with a slope of 0.37 - 0.38 (black dashed lines), while for $El < 1$, the data follows the Wagner limit (green dashed lines), i.e., a constant with values of 3.65 and 3.42 for silane and glass, respectively. The blue dashed line represents the scaling law derived for the Hertzian limit (see text), while the orange dashed line indicates the scaling law based on neo-Hookean elastic energy formulation (Eq. (7) in the main text). Note that the transition from the Hertzian scaling laws to the theoretical Wagner limit (3.24) occurs at $El = 0.3$ while the experimental Wagner limits at $El = 1$.

### 3.4. Post-impact drop dynamics

The morphology of hydrogels while receding after maximum deformation is shown in Figure 8, where $t = 0$ ms corresponds to the time instant of maximum deformation. Following the stage of maximum deformation, except for the stiffest hydrogel, i.e., PAAm30 ($G$ = 130.94 kPa, $El$ = 14.3, Figure 8(e)), all other hydrogel drops undergo a receding stage but notably fail to detach or bounce from the substrate (Figure 8(a-d)). This rebound suppression is primarily attributed to strong interfacial adhesion to the substrate. While glass is inherently hydrophilic, even silanized hydrophobic surfaces exhibit a significant affinity toward the PAAm polymer networks.[38,39]

The contact foot previously observed to pin in softer hydrogels is now observed in even moderate and highly elastic hydrogels to provide an interfacial anchor (Figs. 8a-d). This anchoring can be attributed to the presence of multiple polymer fragments within the contact foot, which remain adsorbed to the substrate and exert a radial resistive force on the bulk drop. This behavior is further supported by the fact that impacted drops left overnight to dry due to evaporation reveal distinct radial lines of PAAm polymer around the bulk gel network (Figure S3, Supplementary Information). Furthermore, FTIR analysis performed on a gold substrate confirms the presence of characteristic PAAm peaks within this residual footprint, verifying the chemical nature of the adhesive layer (Figure S4, Supplementary Information).

To elaborate, we note that, during impact, as the elastic energy is converted back into vertical inertia and the bulk of the drop is pulled upward away from the surface, the pinned polymer network resists this upward momentum, leading to stretching of the contact foot. This interplay of inertial pulling of the bulk, interfacial anchoring of the polymer network, and surface energy minimization of the liquid leads to the formation of distinct circumferential ridges (see last image sequence in Figure 8(a-d)). We observe that as the elasticity increases, the characteristic foot height and the prominence of these ridges decrease, yet the morphological feature persists. This phenomenon is a morphological parallel to the impact of viscoelastic liquid drops on superhydrophobic surfaces, where rebound suppression often results in the formation of a thinning liquid filament.[40,41] In our system, the transition from a liquid-like to a soft solid-like response replaces the singular filament with a deforming contact foot characterized by multiple ridges. These observations are consistent with the ridge patterns reported by Mora *et al*. in their study of vertically hanging gel columns, suggesting that such instabilities are a fundamental characteristic of low-modulus solids subjected to high-strain retraction at a pinned boundary.[42] However, for the impact of the stiffest hydrogel, i.e., PAAm30 ($G$ = 130.94 kPa, $El$ = 14.3, Figure 8(e)), high elasticity, short contact time, and small contact foot lead to reduced adhesion, which can be easily overcome by the elastic energy of the drop. Thus, pinning is not observed, and the drop rebounds.

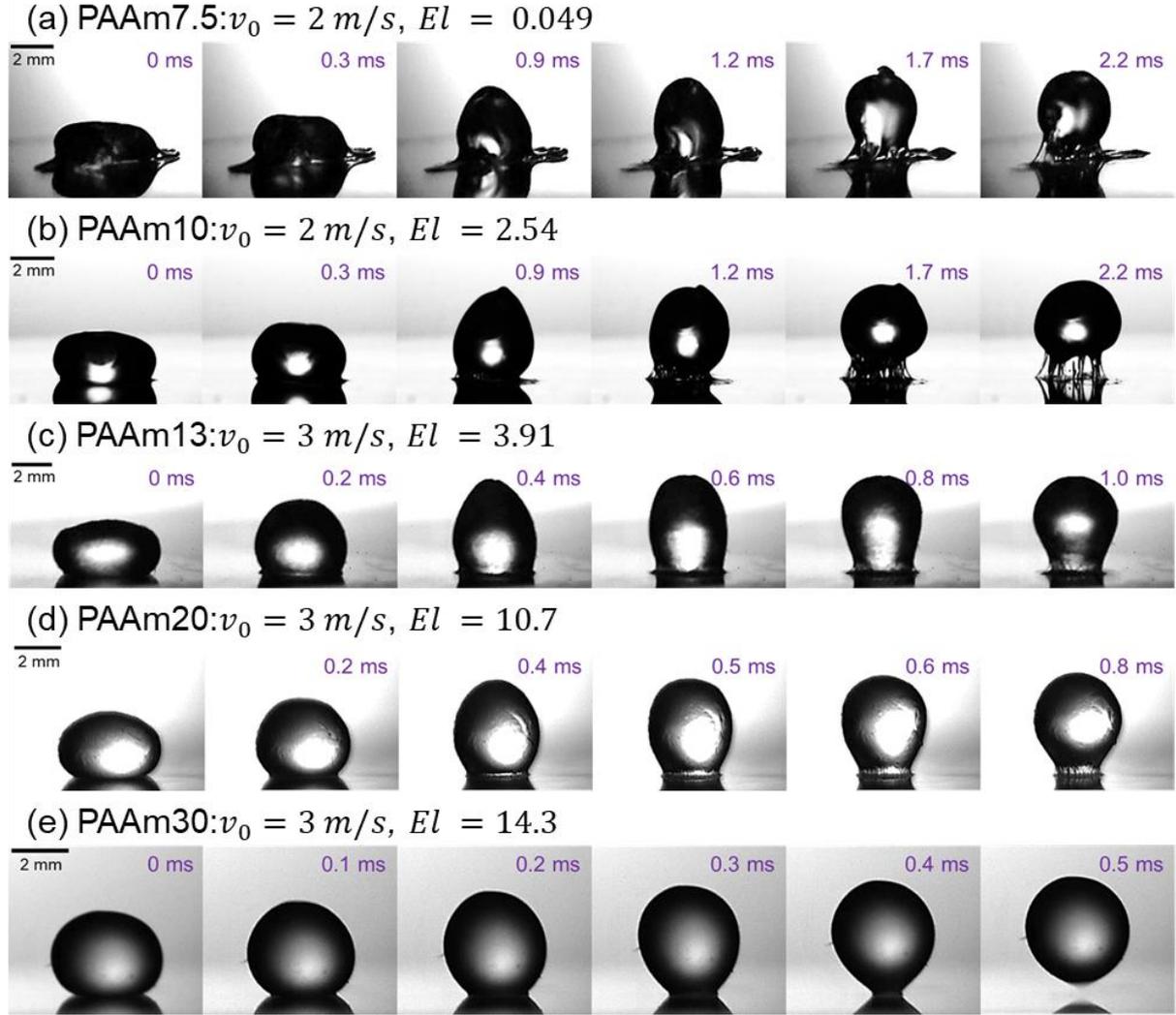

**Figure 8: Post-impact receding dynamics and morphological evolution.** (a–e) High-speed image sequences capturing the receding stage for PAAm hydrogel drops across a wide range of elastic numbers ($El = 0.017 - 14.3$). Note that $t = 0$ represents the moment of maximum spreading diameter post-impact. Further, note that except for PAAm30, the PAAm drops fail to rebound despite significant vertical translation due to the anchoring of the contact foot to the substrate. (a–d) As elastic energy is converted into upward inertia, the pinned base undergoes longitudinal stretching, leading to the formation of circumferential ridges due to the competition between inertial pulling and surface energy minimization. These ridges become less prominent as the gel stiffness increases. (e) Transition to rebound for the stiffest hydrogel (PAAm30), where the high elasticity provides sufficient restorative force to overcome interfacial adhesion, leading to complete detachment. Scale bars represent 2 mm.

## 4. Conclusion

In summary, we investigated the impact dynamics of spherical polyacrylamide (PAAm) hydrogel drops on hydrophilic and hydrophobic substrates across a broad parameter space relevant to bioprinting. By systematically varying the hydrogel shear moduli ($G \approx 0.05 - 131$ kPa) and impact velocities ($We \approx 30 - 270$), we successfully mapped the transition from a liquid-dominated regime to a solid-dominated regime with the characteristic elastic number varying over five decades ($El \approx 0.006 - 128$). Utilizing synchronized high-speed imaging and piezoelectric force sensing, we obtained a simultaneous resolution of transient spreading morphology and impact force variation. This integrated approach provides a comprehensive and internally consistent characterization of soft gel impact, bridging the gap between fluid mechanics and neo-Hookean elastic deformation.

Our results reveal that the elastic number is the primary control parameter governing hydrogel impact behavior. At low elastic numbers ($El < 1$), hydrogels exhibit a distinctly hybrid response: a liquid-rich 'contact foot' is expelled from the network and spreads like a viscous fluid, while the bulk of the drop undergoes viscoelastic pinning and attains a pancake-shaped morphology at maximum deformation. In this regime, a significant fraction of the initial kinetic energy is dissipated through poroelastic flow and contact-line pinning, causing the spreading behavior that deviates from both the liquid and purely elastic-solid drop limit. Conversely, as elasticity increases ($El > 1$), the contribution of contact foot spreading rapidly diminishes, and the deformation of the drop is well described by a neo-Hookean elastic solid undergoing finite strain. In this elasticity-dominated regime, an energy balance of the initial kinetic energy and the final stored elastic strain energy accurately predicts the maximum spreading diameter, independent of substrate wettability. This establishes a clear quantitative transition from dissipation-mediated to elasticity-controlled impact mechanics.

Direct measurements of the transient impact force further corroborate this transition. For $El < 1$, the normalized peak force collapses to a constant value consistent with Wagner's limit for liquid drops,[22–24] reflecting a viscous-dominant momentum transfer. In contrast, for $El > 1$, the peak force increases systematically with elasticity and follows a power-law scaling $F^* \sim El^{0.38}$, in close agreement with both Hertzian[25] and neo-Hookean predictions.

Importantly, this force scaling is independent of surface wettability, indicating that bulk elasticity dominates peak force transmission. From an engineering perspective, this scaling law offers a vital predictive framework for estimating transient mechanical loads on sensitive substrates—such as cell-laden scaffolds, functional membranes, or pre-deposited hydrogel layers—thereby preventing structural damage or delamination in high-speed bioprinting workflows.[26,27]

Post-impact retraction dynamics highlight the critical role of interfacial adhesion and polymer–surface interactions. Across nearly the entire parameter space explored (except PAAm30, $G = 130.94$ kPa), rebound is strongly suppressed on both hydrophilic and hydrophobic substrates. We attribute this to the residual polymer chains within the 'contact foot,' which remain adsorbed to the surface and anchor the retreating gel network. This interfacial anchoring leads to localized stretching instabilities, manifesting as the distinct circumferential ridges observed in our high-speed sequences. It is only when elastic restoring forces significantly exceed the work of adhesion that complete rebound occurs. These observations demonstrate that adhesion dominates the macroscopic outcome of soft hydrogel impact.[40–42]

Overall, this study establishes a unified physical framework for understanding the impact of elastic drops that connect spreading morphology, force transmission, energy dissipation, and rebound suppression through a single governing parameter. By bridging classical fluid mechanics and neo-Hookean elastic solid contact mechanics, our results provide quantitative guidelines for controlling deposition footprint and mechanical loading in hydrogel-based manufacturing. Additionally, these insights are directly relevant to high-resolution bioprinting for controlled deposition of biomaterial inks, where precise management of both geometry and mechanical strength is essential for optimal performance.[43]

## CRediT authorship contribution statement

**Akash Chowdhury:** Writing – review & editing, Writing – original draft, Data curation, Validation, Methodology, Investigation. **Surjyasish Mitra:** Writing – review & editing, Visualization, Validation, Investigation. **Sushanta K. Mitra:** Conceptualization, Writing – review & editing, Supervision, Project administration, Funding acquisition.

## Declaration of Competing Interest

None

## Acknowledgements

S. K. M. acknowledges the support of the Discovery Grant (NSERC, RGPIN-2024-03729). The authors also acknowledge the use of OpenAI's ChatGPT platform for minor paraphrasing and proof-reading parts of the manuscript.

## Appendix A. Supplementary Data

All data that support the findings of this study are included within the article. The data supporting this article have been included as part of the supplementary information (SI). Supplementary information is available.